\documentclass[journal=jacsat,manuscript=article]{achemso}
\usepackage{chemformula} 
\usepackage[T1]{fontenc} 

\author{Inna Krasnokutska}
\email{inna@xanadu.ai}
\affiliation[rmit]{Quantum Photonics Laboratory and Centre for Quantum Computation and Communication Technology, School of Engineering, RMIT University, Melbourne, Victoria 3000, Australia}
\alsoaffiliation[new adress]{Current address: Xanadu Quantum Technologies Inc., 2902-777 Bay St, Toronto, ON, M5G 2C8}
\author{Jean-Luc J. Tambasco}
\altaffiliation{Current address: Cisco Systems, Inc., 7540 Windsor Drive, Suite 412, Allentown, PA 18195, USA}
\author{Alberto Peruzzo}
\affiliation[rmit]
{Quantum Photonics Laboratory and Centre for Quantum Computation and Communication Technology, School of Engineering, RMIT University, Melbourne, Victoria 3000, Australia}

\title{Submicron domain engineering in periodically
poled lithium niobate on insulator}
\begin{document}

\begin{abstract}

Lithium niobate on insulator (LNOI) is extensively researched for potential applications in telecommunication, sensing and quantum technology. Low-loss waveguides in LNOI have been demonstrated in recent years, with further works investigating their nonlinear optical properties via electric field poling. The fabrication imperfections and minimum feature size of inverted ferroelectric domains currently limit the efficiency and application of nonlinear processes in LNOI. This work looks into focused ion beam poling and demonstrates its potential to overcome the limitations of electric field poling. We numerically investigate the susceptibility of quasi-phase-matching condition to the fabrication flaws in poling periods and waveguide geometry showing its extreme sensitivity. We demonstrate the precise fabrication of poling periods down to 200\,nm consistent over 3$\,$mm length. Submicron poling of low loss LNOI waveguides will unlock unprecedentedly efficient nonlinear interactions by the use of uncomplicated LNOI components and enable the realization of mirrorless optical parametric oscillators.

\end{abstract}

\section{Introduction}

Lithium niobate on insulator (LNOI) is a promising photonic platform that combines the material properties of lithium niobate (LN) with high index contrast photonic circuits \cite{zhu2021integrated,Krasnokutska:18,Zhang:17}. LNOI has demonstrated unique applications in telecommunication \cite{Mercante:18, wang2018integrated, krasnokutska2019tunable}, sensing \cite{YAO2020103082} and quantum technology\cite{doi:10.1063/1.5054865, PhysRevLett.124.163603}; however, the full potential of optical nonlinearities in LNOI is yet to be fully explored. 

Highly efficient nonlinear processes in LNOI rely on three-wave mixing by use of the $\chi ^2$ optical susceptibility, which is naturally inefficient due to material dispersion, and results in phase mismatch between the interacting waves. Quasi-phase-matching (QPM), conventionally implemented through electric field poling (EFP), is used to fulfil the phase matching conditions by creating periodic perturbations in the nonlinear susceptibility, satisfying the condition for efficient three-wave mixing \cite{myers1995quasi, Tambasco:16}. Periodically poled $X$-cut and $Z$-cut LNOI (PPLNOI) photonic components have been recently demonstrated \cite{Wang:18, Rao:19,Lu:19,Chen:20}, however, their performance is restricted by technological challenges. The size of QPM periods that can be reliably fabricated via the EFP technique is limited to $\sim$2-3\,$\mu m$, setting nonlinear interactions to only forward harmonic generation \cite{Zhu:21, doi:10.1080/23746149.2021.1889402}. Furthermore, achieving homogeneous, errorless PPLNOI structures with a 50 $\%$ duty cycle is challenging for small QPM periods since high precision control of EFP parameters and high-quality fabrication of poling electrodes are required. Often, the waveguide geometry is engineered to fit into the fabrication conditions compromising the device performance. Periodic inversion of ferroelectric domains by use of piezo force microscopy (PFM) \cite{Hao:20, Gainutdinov, doi:10.1063/1.5054865}, electron beam and focused ion beam (FIB) have been also investigated \cite{LI:2006, shur}. These techniques are often incompatible with low loss waveguide fabrication and may result in unstable poling domain inversion or lack of scalability.

Submicron domain inversion is needed to enable backward optical harmonic generation that makes use of implicit cavity effects to drastically enhance nonlinear interactions without the need for integrated photonic cavities like microring resonators and Bragg grating cavities. Higher-order QPM backward harmonic generation for bulk lithium niobate and low index contrast waveguides has already been reported, highlighting the challenges of the fabrication of submicron domains \cite{Kang,Luo:20}. The demonstration of the first order backward second-harmonic generation in the telecom range would require a QPM period $\sim$200\,nm, which is technologically challenging via current poling technologies.

We demonstrate a one-step, reliable and high yield fabrication process of $Z$-cut PPLNOI via FIB poling compatible with any waveguide geometry. We emphasize the importance of the advanced ferroelectric domain engineering in LNOI by studying the QPM periods and the fabrication tolerances required to achieve forward-forward (FF), forward-backward (FB) and backward-backward (BB) SHG in single mode LNOI waveguides. We then evaluate the quality of the obtained poling pattern, analyze the fabrication tolerances and the consistency of poling throughout the LNOI film. We demonstrate periodically poled $Z$-cut LNOI films with less than 3\,$\%$ deviation from the design period and confirm that the FIB poling technique does not adversely affect the underlying LNOI ridge waveguide. Lastly, we report submicron PPLNOI with a period $\Lambda \sim$200\,nm and duty cycle $D \sim$33\,$\%$. 

\begin{figure}
\includegraphics[width=0.8\linewidth]{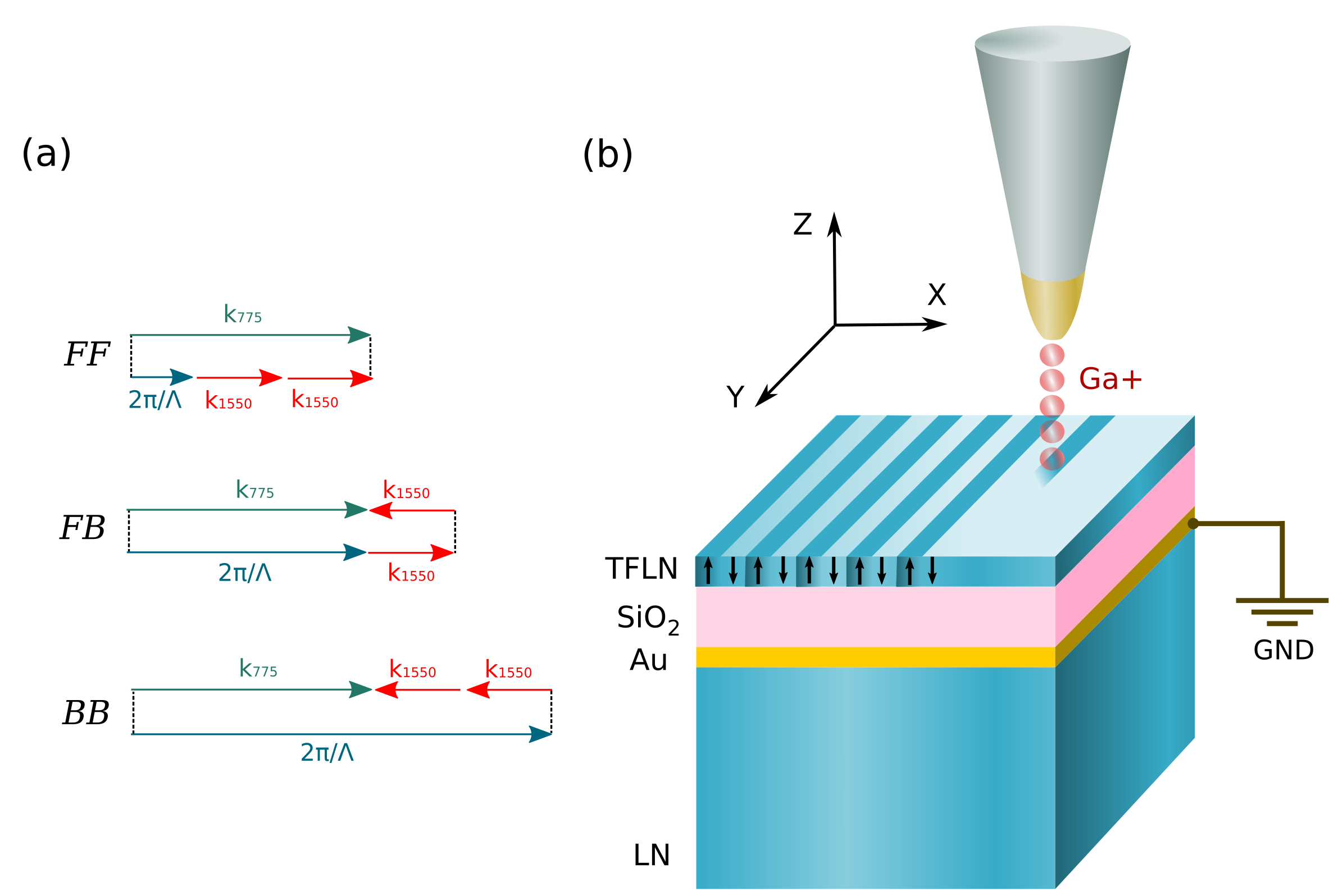}
  \caption{ (a) Configurations of quasi phase matched SHG: forward-forward (FF), forward-backward (FB) and backward-backward (BB). (b) Schematic of the formation of poled regions in lithium niobate on insulator.}
  \label{fgr:fibsetup}
\end{figure}

\section{Design and analysis}

We consider three configurations of SHG shown in Fig. \ref{fgr:fibsetup}(a) to highlight the importance of advanced ferroelectric domain engineering. 

We first start with forward-forward (FF) SHG. All three interacting waves are propagating in the same direction; two photons with $\lambda = 1550$\,nm and a wavevector $k_{1550}$ (highlighted in red) may generate one photon with $\lambda = 775$\,nm and wave vector $k_{775}$ (highlighted in green). The QPM condition is satisfied by the mismatch compensation vector $2\pi/\Lambda$, where $\Lambda$ is a poling period.

In the case of forward-backward (FB) SHG, the QPM period decreases compared to the FF case. A pump and generated signal are propagating in the forward direction; the other pump photon is counterpropagating as shown in Fig. \ref{fgr:fibsetup}(a). The poling period further reduces for backward-backward (BB) SHG, where both pumps are forward propagating and the generated signal is counter-propagating. The perturbations in the second-order susceptibility due to a particular period in ferroelectric domain inversion create effects similar to the nonlinear media, that is placed inside of an optical cavity, but with the absence of the physical mirrors.

The FF SHG poling period for bulk lithium niobate is about $\sim18 \,\mu$m. In a highly dispersive media, like a single mode LNOI waveguide, the QPM period reduces almost by a factor of 3 leading to further complications in satisfying phase-matching condition. 

The required poling periods can be calculated using a fully-vectorial mode solver and the QPM conditions in Fig. \ref{fgr:fibsetup}. The calculations of the effective index are used to reconstruct the QPM condition for a particular waveguide geometry. 

The simulated poling periods for FF, FB and BB SHGs are presented in Tab. \ref{tbl:periods}. The simulations were performed for the waveguides with 1\,$\mu$m of bottom width and 350\,nm of etching depth. The total film thickness is 500\,nm. We consider etching profiles of $X$-cut LNOI with the sidewall angle of $70^\circ$ and $Z$-cut with $75^\circ$ reported in [\cite{Krasnokutska:s}]. The waveguides are cladded with a 3\,$\mu$m thick SiO$_{2}$ layer. We also perform simulations of the poling period for unetched LNOI thin films in Tab. \ref{tbl:periods}.

\begin{table}
  \caption{Simulated QPM periods for 1\,$\mu$m waveguide and thin 500\,nm thick $X$-cut and $Z$-cut LNOI films. The SHG pump wavelength is 1550\,nm.}
  \label{tbl:periods}
  \begin{tabular}{llll}
    \hline
    Device  & FF, $\mu$m & FB, $\mu$m & BB, $\mu$m \\
    \hline 
    X-cut LNOI waveguide & 3.460 & 0.370 & 0.196   \\
    Z-cut LNOI waveguide  & 2.660 & 0.375 & 0.200 \\
    Z-cut LNOI unetched (glass clad)  & 3.220  & 0.372 & 0.197 \\
    X-cut LNOI unetched (glass clad)  & 4.630  & 0.368 & 0.190 \\
  \hline
  \end{tabular}
\end{table}

\begin{figure}
\includegraphics[width=1.0\linewidth]{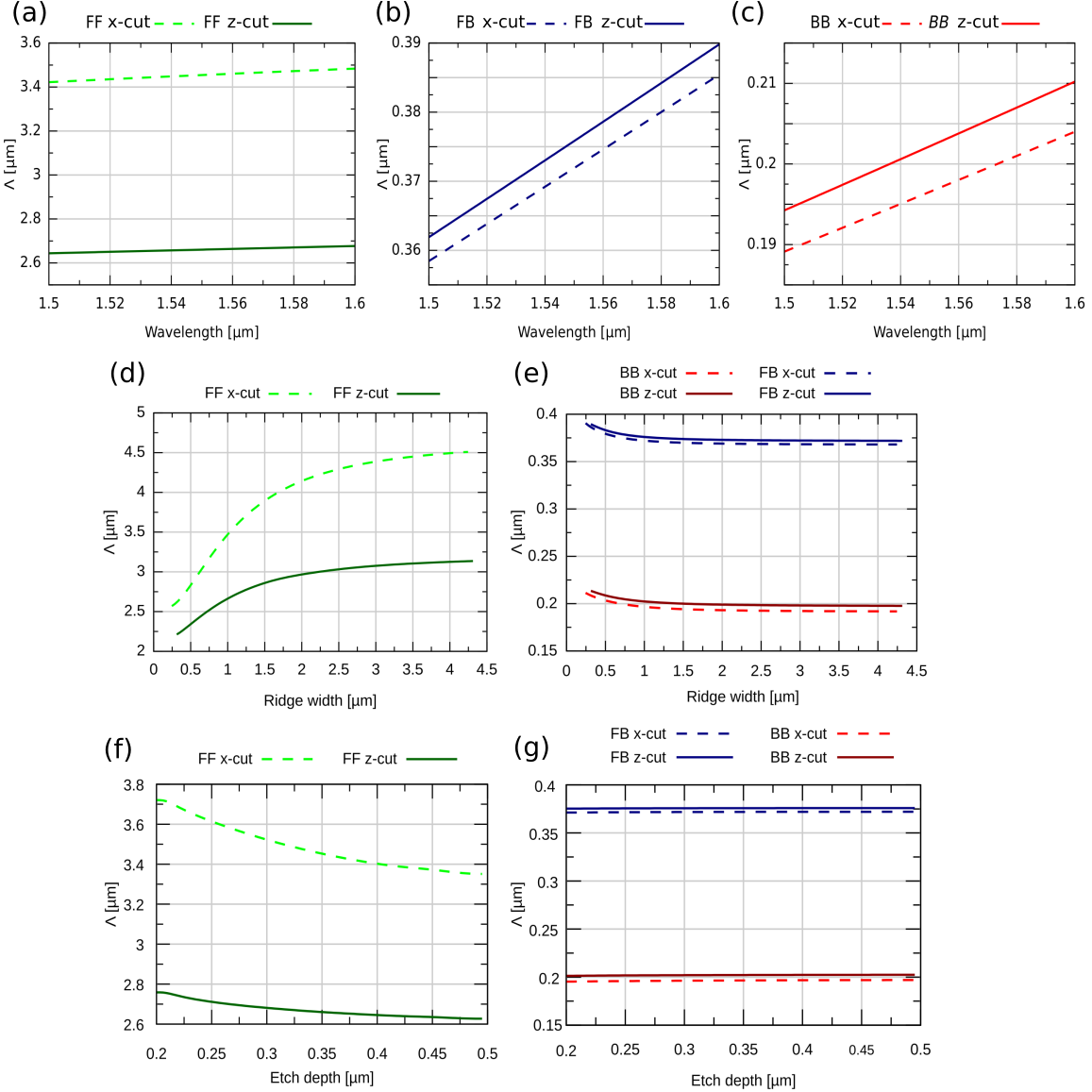}
  \caption{Sensitivity of QPM condition to poling period variation: (a) forward-forward SHG (FF), (b) forward-backward SHG (FB) and (c) backward-backward SHG (BB) for both $X$- and $Z$-cut LN. Waveguide fabrication tolerances: (d) and (e) influence of variations in the ridge width on the QPM period for forward-forward and cavity enhanced (forward-backward and backward-backward) second harmonic generations respectively at 1550\,nm pump wavelength; (f) and (g) influence of variations in the etch depth on the QPM period for forward-forward and cavity enhanced (forward-backward and backward-backward) second harmonic generations respectively at 1550\,nm pump wavelength.}
  \label{fgr:period}
\end{figure}

The QPM condition is sensitive to the waveguide geometry and poling period. Small variations in the poling period lead to a drastic shift in the SHG pump wavelength as shown in Fig. \ref{fgr:period}. In the case of FF SHG, a 40\,nm deviation from the designed poling period may lead to a QPM shift over both $C$- and $L$-bands (1530-1565\,nm and 1565-1625\,nm respectively) in the Fig. \ref{fgr:period}(a). The FB and BB QPM conditions move over two bands just with 20\,nm change in the poling period as depicted in the Fig. \ref{fgr:period}(b),(c). 

We simulate the sensitivity of the poling period to the change in the waveguide width and the etching depth (Fig. \ref{fgr:period}). We first investigate poling periods for $X$-cut and $Z$-cut LNOI waveguides with the ridge widths 0.2-4.5\,$\mu$m, etching depth 350\,nm and covered with the 3$\mu$m thick SiO$_2$layer (Fig. \ref{fgr:period}(d),(e)). The QPM period is changing rapidly for C-band singlemode and slightly multimode LNOI waveguides (0.7-2\,$\mu$m bottom waveguide widths). QPM condition is less sensitive to the waveguide geometry for highly multimode waveguides. Poling period becomes insensitive to the widths variations for 4\,$\mu$ width waveguides corresponding to the periods of the unpatterned thin film LNOI (Tab. \ref{tbl:periods}). The FB and BB configurations are less sensitive to the waveguide geometry variations and the change in the poling period with the waveguide width are represented by the mostly flat curve in Fig. \ref{fgr:period}(e).

We then investigate the variations in the etching depths 0.2-0.5\,$\mu$m of the $X$-cut and $Z$-cut LNOI waveguides with the bottom ridge width 1\,$\mu$m covered with the 3\,$\mu$m SiO$_2$ layer (Fig. \ref{fgr:period}(f),(g)). $X$-cut LNOI is more sensitive to the variations in the waveguide geometry. It can be explained by the slightly better etching profile of $Z$-cut LNOI waveguides. Just 50\,nm deviation in the designed etching depth can cause the 40\,nm and 100\,nm change in the poling period for $Z$-cut and $X$-cut LNOI waveguides respectively, as shown in Fig. \ref{fgr:period}(f). The FF and FB generations as in the case with the ridge width variations are almost insensitive to the fabrication imperfections (Fig. \ref{fgr:period}(g)). 

\section{Experimental methodology}

We first perform FIB patterning of untreated $+Z$-cut LNOI surface. We start with 500\,nm LN thin film placed on 2\,$\mu$m SiO$_2$ layer fabricated via smart cut technique and supported by a 500\,$\mu$m thick bulk LN substrate as shown in Fig \ref{fgr:fibsetup}. The material stack contains a Cr/Au/Cr electrode layer underneath the SiO$_2$ bottom cladding, which is electrically grounded to the FIB loading metal holder. We use a 35\,kV voltage Ga+ FIB with the 40-60\,$\mu$m apertures to achieve step size resolution of 6-10\,nm for the chosen material.

FIB patterning introduces surface roughness as a result of thin film milling and consequent material redeposition. The amount of roughness depends on the used FIB aperture and varies 5-10\,nm RMS in our case based on the calibration runs.

The post-patterning surface roughness can be reduced by the spin coating of 50\,nm thick PMMA layer over the LNOI surface. The presence of the resist layer also contributes to the homogeneous growth of poling domains \cite{LI:2006}. The PMMA is removed in the $2\%$ TMAH solution after patterning.

We sweep the FF poling periods in the range 2.3 - 2.98\,$\mu$m with 40\,nm step to take into account the sensitivity of QPM condition to the fabrication imperfections in Fig. \ref{fgr:period}. The theoretically predicted poling period from Tab. \ref{tbl:periods} is located in the middle of the sweep range. The target $D$ is 50\,$\%$. As preliminary fabrication test show that poling domains slightly broaden, introducing error to the duty cycle $\sim$3-5\,$\%$, we design the duty cycle to be 45\,$\%$ to compensate for the possible overpoling. 

FB and BB poling periods are extremely challenging to achieve, but the QPM conditions are less sensitive to the fabrication imperfections and the waveguide geometry variations. We fabricate multiple periods around the designed values in Tab. \ref{tbl:periods}. We perform the prefabrication tests by fabricating the same poling periods with different duty cycles in the range of 10-45\,$\%$. As poling domains broaden and merge quickly for $D$=35-45\,$\%$ due to strong proximity effects, we keep the duty cycle $\sim$20-30\,$\%$ to be close as possible to the target $D\sim$50\,$\%$, but compensate for domain broadening.

The length of each poling pattern is 1 and 3\,mm. That allows reducing errors related to overlaying the poling pattern over the fabricated waveguide and also facilitates the PFM measurements.

\begin{figure}
\includegraphics[width=0.7\linewidth]{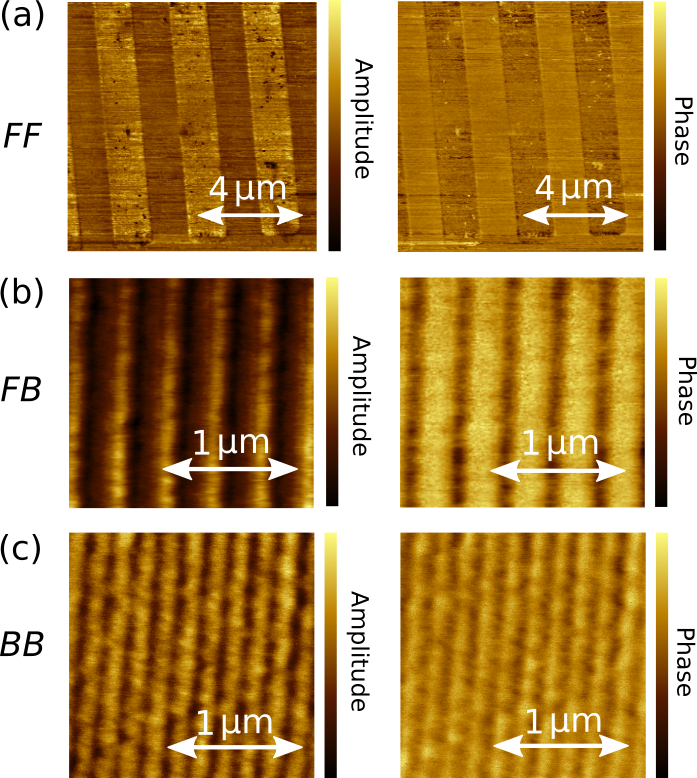}
  \caption{PFM phase and amplitude of PPLNOI: (a) with a period $\sim$2.82\,$\mu$m and $\sim$47\,$\%$ duty cycle for forward-forward configuration; (b) with a period $\sim$380\,nm and $\sim$40\,$\%$ duty cycle for forward-backward configuration; (c) with a period $\sim$200\,nm and $\sim$35\,$\%$ duty cycle for backward-backward configuration. }
  \label{fgr:pfm}
\end{figure}

We first characterize the poling quality using piezoresponse force microscopy (PFM). Poling periods and their duty cycles can be assessed with submicron resolution. We use a low voltage setting of Cypher AFM by Asylum Research to prevent polarization back switching that can introduce errors to the poling pattern and avoid possible damage to the sample by reaching a dielectric breakdown. The displacement of a conductive tip is directly related to the polarization direction of $d_{33}$ below the tip. Poling cannot be seen under an optical microscope after the PMMA layer is been removed. We mark each poling pattern with different $\Lambda$ before patterning and investigate at different places under PFM to make sure that the poling is consistent over the whole pattern length. We can observe periodic structures with $\Lambda\sim$2.82\,$\mu$m as designed and  $D$=47\,$\%$ with a deviation by $\sim$2\,$\%$ for both PFM phase and amplitude in Fig. \ref{fgr:pfm}(a). These results are consistent for each FF poling period, and no errors are observed in the poling patterns.

Fig. \ref{fgr:pfm}(b) and (c) demonstrate 0.38\,$\mu$m period required for FB-QPM condition  0.2\,$\mu$m period required for BB-QPM condition correspondingly. The poling patterns exhibit deviation of 10-15\,$\%$ from the designed poling period as expected due to domain broadening, which is off by 10-15\,$\%$ from the target duty cycle $D=50\,\%$.  It should be noted that no degradation is observed in the poling pattern over time.

\begin{figure}
\includegraphics[width=0.85\linewidth]{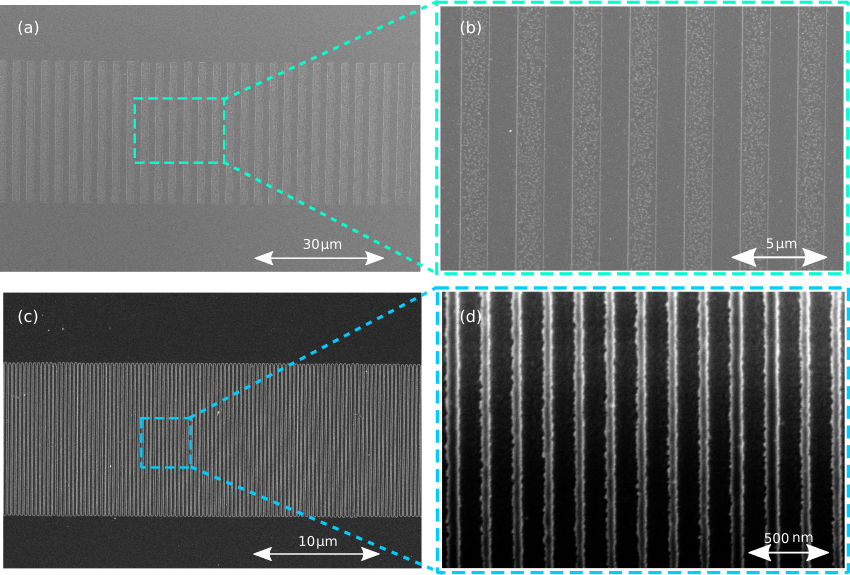}
  \caption{Scanning electron microscopy pictures of poling pattern revealed after selective HF etching indicating good uniformity along the poling length: (a) periodically poled domains with $\Lambda\sim3\,\mu$m and $D\sim48\,\%$ with (b) a close up view; (c) periodically poled domains with $\Lambda\sim$198\,nm and $D\sim33\,\%$ with (d) a close up view.}
  \label{fgr:Ultrasmall}
\end{figure}

We then investigate the quality of obtained poling pattern by etching samples in the 10\,$\%$ diluted HF. Due to the difference in the chemical etching rate of $+Z$ and $-Z$ faces the periodic domain structuring can be seen under a scanning electron microscope in Fig. \ref{fgr:Ultrasmall}. We verify that the poling pattern is consistent over the whole length, and constant duty cycle and no broadening in the poling period were observed.

To verify the consistency of the poling domains throughout the whole film thickness, we argon mill the poled film leaving 100\,nm of residual LN film. Mechanical etching is almost insensitive to the LN crystal orientation and it can be used for evenly and precisely thinning the film away. We again perform the etching of the rest of LN film and check the poled pattern under SEM and measure the height of formed structures by using a profilometer. The poling domains are preserved through the LN film and no distortion in the pattern are found.

\begin{figure}
\includegraphics[width=1.0\linewidth]{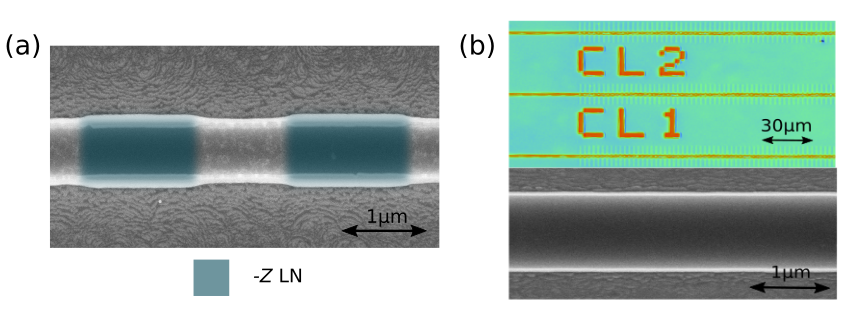}
  \caption{a) False-colour scanning electron microscopy pictures: poled waveguide etched 300\,nm with period of $\sim2.5$\,$\mu$m; (b) Poling pattern visible under optical profilometer on PMMA layer over an etched waveguide and scanning electron microscopy picture of the waveguide after PMMA layer is removed. No corrugations or FIB induced sidewall roughness are observed.}
  \label{fgr:etchedwg}
\end{figure}

FIB poling allows performing the poling either before or after waveguide fabrication. The fabrication of the PPLN pattern before the waveguide fabrication results in the periodic corrugation along the waveguide due to the difference in the mechanical and chemical etching rate of $+Z$ and $-Z$ faces. This may result in increased propagation loss due to sidewall scattering and it is particularly pronounced during the etching in fluorine-based plasma. We investigate this effect by first poling the thin film with desired FF-SHG period and then fabricating the single mode waveguide over the poling pattern. Fig. \ref{fgr:etchedwg}(a) shows the top view of the $\sim$1\,$\mu$m width waveguide that has been etched down to 300\,nm. We observe the distinct $>$15\,$\%$ periodic variations in the waveguide width due to the difference in etch rates between the original and inverted crystalline axes. We then first etch the waveguide and perform poling afterwards on another sample.  Fig. \ref{fgr:etchedwg}(b) shows realized poling pattern over PMMA covered waveguide under optical profilometer. PMMA layer is again used to reduce the surface roughness and promote homogeneous growth of poling domains. The bottom inset in Fig. \ref{fgr:etchedwg}(b) is a scanning electron microscopy picture of the poled waveguide after striping the PMMA protective layer.

Similar variations in the waveguide dimensions have been observed for $X$-cut and $Z$-cut PPLNOI fabricated via EFP techniques \cite{Wang:18}. The EFP and PFM poling techniques are challenging to implement after waveguide fabrication making the corrugations introduced by the domain reversal unavoidable. FIB poling overcomes this limitation in $Z$-cut LNOI as it does not require the fabrication of the additional electrodes onto the waveguide.

\section{Conclusion}

We investigated the sensitivity of the LNOI QPM condition to the fabrication variations in the waveguide geometry and poling period. The FIB poling technique enables the submicron period fabrication required for the realization of cavity-like nonlinear processes in LNOI. We have realized poling periods down to $\sim$200\,nm with $\sim$35\,$\%$ duty cycle over 3\,mm waveguide length. The poling pattern is verified via PFM to have $\sim$100\,$\%$ yield over the full length and through the film thickness. FIB poling allows fast domain patterning compared to the PFM technique. It can be performed before or after waveguide fabrication without introducing additional roughness, although poling before ridge etching will lead to waveguide width corrugations. The FIB poling technique may unlock the full potential of the nonlinear processes in LNOI without compromising performance and waveguide geometry.

\section*{Acknowledgments}
A.P. acknowledges a RMIT University Vice-Chancellor's Senior Research Fellowship; ARC DECRA Fellowship (No: DE140101700), and Google Faculty Research Award. This work was supported by the Australian Government through the Australian Research Council under the Centre of Excellence scheme (No: CE170100012, CE170100026).
This work was performed in part at the Melbourne Centre for Nanofabrication in the Nanolab at Swinburne University of Technology. The authors acknowledge the facilities, and the scientific and technical assistance, of the Australian Microscopy \& Microanalysis Research Facility at RMIT University.


\begin{mcitethebibliography}{25}
\providecommand*\natexlab[1]{#1}
\providecommand*\mciteSetBstSublistMode[1]{}
\providecommand*\mciteSetBstMaxWidthForm[2]{}
\providecommand*\mciteBstWouldAddEndPuncttrue
  {\def\EndOfBibitem{\unskip.}}
\providecommand*\mciteBstWouldAddEndPunctfalse
  {\let\EndOfBibitem\relax}
\providecommand*\mciteSetBstMidEndSepPunct[3]{}
\providecommand*\mciteSetBstSublistLabelBeginEnd[3]{}
\providecommand*\EndOfBibitem{}
\mciteSetBstSublistMode{f}
\mciteSetBstMaxWidthForm{subitem}{(\alph{mcitesubitemcount})}
\mciteSetBstSublistLabelBeginEnd
  {\mcitemaxwidthsubitemform\space}
  {\relax}
  {\relax}

\bibitem[Zhu \latin{et~al.}(2021)Zhu, Shao, Yu, Cheng, Desiatov, Xin, Hu,
  Holzgrafe, Ghosh, Shams-Ansari, Puma, Sinclair, Reimer, Zhang, and
  Lon\v{c}ar]{zhu2021integrated}
Zhu,~D.; Shao,~L.; Yu,~M.; Cheng,~R.; Desiatov,~B.; Xin,~C.~J.; Hu,~Y.;
  Holzgrafe,~J.; Ghosh,~S.; Shams-Ansari,~A.; Puma,~E.; Sinclair,~N.;
  Reimer,~C.; Zhang,~M.; Lon\v{c}ar,~M. Integrated photonics on thin-film
  lithium niobate. 2021\relax
\mciteBstWouldAddEndPuncttrue
\mciteSetBstMidEndSepPunct{\mcitedefaultmidpunct}
{\mcitedefaultendpunct}{\mcitedefaultseppunct}\relax
\EndOfBibitem
\bibitem[Krasnokutska \latin{et~al.}(2018)Krasnokutska, Tambasco, Li, and
  Peruzzo]{Krasnokutska:18}
Krasnokutska,~I.; Tambasco,~J.-L.~J.; Li,~X.; Peruzzo,~A. \emph{Opt. Express}
  \textbf{2018}, \emph{26}, 897--904\relax
\mciteBstWouldAddEndPuncttrue
\mciteSetBstMidEndSepPunct{\mcitedefaultmidpunct}
{\mcitedefaultendpunct}{\mcitedefaultseppunct}\relax
\EndOfBibitem
\bibitem[Zhang \latin{et~al.}(2017)Zhang, Wang, Cheng, Shams-Ansari, and
  Lon\v{c}ar]{Zhang:17}
Zhang,~M.; Wang,~C.; Cheng,~R.; Shams-Ansari,~A.; Lon\v{c}ar,~M. \emph{Optica}
  \textbf{2017}, \emph{4}, 1536--1537\relax
\mciteBstWouldAddEndPuncttrue
\mciteSetBstMidEndSepPunct{\mcitedefaultmidpunct}
{\mcitedefaultendpunct}{\mcitedefaultseppunct}\relax
\EndOfBibitem
\bibitem[Mercante \latin{et~al.}(2018)Mercante, Shi, Yao, Xie, Weikle, and
  Prather]{Mercante:18}
Mercante,~A.~J.; Shi,~S.; Yao,~P.; Xie,~L.; Weikle,~R.~M.; Prather,~D.~W.
  \emph{Opt. Express} \textbf{2018}, \emph{26}, 14810--14816\relax
\mciteBstWouldAddEndPuncttrue
\mciteSetBstMidEndSepPunct{\mcitedefaultmidpunct}
{\mcitedefaultendpunct}{\mcitedefaultseppunct}\relax
\EndOfBibitem
\bibitem[Wang \latin{et~al.}(2018)Wang, Zhang, Chen, Bertrand, Shams-Ansari,
  Chandrasekhar, Winzer, and Lon{\v{c}}ar]{wang2018integrated}
Wang,~C.; Zhang,~M.; Chen,~X.; Bertrand,~M.; Shams-Ansari,~A.;
  Chandrasekhar,~S.; Winzer,~P.; Lon{\v{c}}ar,~M. \emph{Nature} \textbf{2018},
  \emph{562}, 101--104\relax
\mciteBstWouldAddEndPuncttrue
\mciteSetBstMidEndSepPunct{\mcitedefaultmidpunct}
{\mcitedefaultendpunct}{\mcitedefaultseppunct}\relax
\EndOfBibitem
\bibitem[Krasnokutska \latin{et~al.}(2019)Krasnokutska, Tambasco, and
  Peruzzo]{krasnokutska2019tunable}
Krasnokutska,~I.; Tambasco,~J.-L.~J.; Peruzzo,~A. \emph{Scientific reports}
  \textbf{2019}, \emph{9}, 1--7\relax
\mciteBstWouldAddEndPuncttrue
\mciteSetBstMidEndSepPunct{\mcitedefaultmidpunct}
{\mcitedefaultendpunct}{\mcitedefaultseppunct}\relax
\EndOfBibitem
\bibitem[Yao \latin{et~al.}(2020)Yao, Liu, Zhang, Liu, and Liu]{YAO2020103082}
Yao,~Y.; Liu,~B.; Zhang,~H.; Liu,~H.; Liu,~J. \emph{Results in Physics}
  \textbf{2020}, \emph{17}, 103082\relax
\mciteBstWouldAddEndPuncttrue
\mciteSetBstMidEndSepPunct{\mcitedefaultmidpunct}
{\mcitedefaultendpunct}{\mcitedefaultseppunct}\relax
\EndOfBibitem
\bibitem[Aghaeimeibodi \latin{et~al.}(2018)Aghaeimeibodi, Desiatov, Kim, Lee,
  Buyukkaya, Karasahin, Richardson, Leavitt, Lončar, and
  Waks]{doi:10.1063/1.5054865}
Aghaeimeibodi,~S.; Desiatov,~B.; Kim,~J.-H.; Lee,~C.-M.; Buyukkaya,~M.~A.;
  Karasahin,~A.; Richardson,~C. J.~K.; Leavitt,~R.~P.; Lončar,~M.; Waks,~E.
  \emph{Applied Physics Letters} \textbf{2018}, \emph{113}, 221102\relax
\mciteBstWouldAddEndPuncttrue
\mciteSetBstMidEndSepPunct{\mcitedefaultmidpunct}
{\mcitedefaultendpunct}{\mcitedefaultseppunct}\relax
\EndOfBibitem
\bibitem[Zhao \latin{et~al.}(2020)Zhao, Ma, R\"using, and
  Mookherjea]{PhysRevLett.124.163603}
Zhao,~J.; Ma,~C.; R\"using,~M.; Mookherjea,~S. \emph{Phys. Rev. Lett.}
  \textbf{2020}, \emph{124}, 163603\relax
\mciteBstWouldAddEndPuncttrue
\mciteSetBstMidEndSepPunct{\mcitedefaultmidpunct}
{\mcitedefaultendpunct}{\mcitedefaultseppunct}\relax
\EndOfBibitem
\bibitem[Myers \latin{et~al.}(1995)Myers, Eckardt, Fejer, Byer, Bosenberg, and
  Pierce]{myers1995quasi}
Myers,~L.~E.; Eckardt,~R.; Fejer,~M.; Byer,~R.; Bosenberg,~W.; Pierce,~J.
  \emph{JOSA B} \textbf{1995}, \emph{12}, 2102--2116\relax
\mciteBstWouldAddEndPuncttrue
\mciteSetBstMidEndSepPunct{\mcitedefaultmidpunct}
{\mcitedefaultendpunct}{\mcitedefaultseppunct}\relax
\EndOfBibitem
\bibitem[Tambasco \latin{et~al.}(2016)Tambasco, Boes, Helt, Steel, and
  Mitchell]{Tambasco:16}
Tambasco,~J.-L.; Boes,~A.; Helt,~L.~G.; Steel,~M.~J.; Mitchell,~A. \emph{Opt.
  Express} \textbf{2016}, \emph{24}, 19616--19626\relax
\mciteBstWouldAddEndPuncttrue
\mciteSetBstMidEndSepPunct{\mcitedefaultmidpunct}
{\mcitedefaultendpunct}{\mcitedefaultseppunct}\relax
\EndOfBibitem
\bibitem[Wang \latin{et~al.}(2018)Wang, Langrock, Marandi, Jankowski, Zhang,
  Desiatov, Fejer, and Lon\v{c}ar]{Wang:18}
Wang,~C.; Langrock,~C.; Marandi,~A.; Jankowski,~M.; Zhang,~M.; Desiatov,~B.;
  Fejer,~M.~M.; Lon\v{c}ar,~M. \emph{Optica} \textbf{2018}, \emph{5},
  1438--1441\relax
\mciteBstWouldAddEndPuncttrue
\mciteSetBstMidEndSepPunct{\mcitedefaultmidpunct}
{\mcitedefaultendpunct}{\mcitedefaultseppunct}\relax
\EndOfBibitem
\bibitem[Rao \latin{et~al.}(2019)Rao, Abdelsalam, Sjaardema, Honardoost,
  Camacho-Gonzalez, and Fathpour]{Rao:19}
Rao,~A.; Abdelsalam,~K.; Sjaardema,~T.; Honardoost,~A.;
  Camacho-Gonzalez,~G.~F.; Fathpour,~S. \emph{Opt. Express} \textbf{2019},
  \emph{27}, 25920--25930\relax
\mciteBstWouldAddEndPuncttrue
\mciteSetBstMidEndSepPunct{\mcitedefaultmidpunct}
{\mcitedefaultendpunct}{\mcitedefaultseppunct}\relax
\EndOfBibitem
\bibitem[Lu \latin{et~al.}(2019)Lu, Surya, Liu, Bruch, Gong, Xu, and
  Tang]{Lu:19}
Lu,~J.; Surya,~J.~B.; Liu,~X.; Bruch,~A.~W.; Gong,~Z.; Xu,~Y.; Tang,~H.~X.
  \emph{Optica} \textbf{2019}, \emph{6}, 1455--1460\relax
\mciteBstWouldAddEndPuncttrue
\mciteSetBstMidEndSepPunct{\mcitedefaultmidpunct}
{\mcitedefaultendpunct}{\mcitedefaultseppunct}\relax
\EndOfBibitem
\bibitem[Chen \latin{et~al.}(2020)Chen, Tang, Ma, Li, Sua, and Huang]{Chen:20}
Chen,~J.-Y.; Tang,~C.; Ma,~Z.-H.; Li,~Z.; Sua,~Y.~M.; Huang,~Y.-P. \emph{Opt.
  Lett.} \textbf{2020}, \emph{45}, 3789--3792\relax
\mciteBstWouldAddEndPuncttrue
\mciteSetBstMidEndSepPunct{\mcitedefaultmidpunct}
{\mcitedefaultendpunct}{\mcitedefaultseppunct}\relax
\EndOfBibitem
\bibitem[Zhu \latin{et~al.}(2021)Zhu, Shao, Yu, Cheng, Desiatov, Xin, Hu,
  Holzgrafe, Ghosh, Shams-Ansari, Puma, Sinclair, Reimer, Zhang, and
  Lon\v{c}ar]{Zhu:21}
Zhu,~D.; Shao,~L.; Yu,~M.; Cheng,~R.; Desiatov,~B.; Xin,~C.~J.; Hu,~Y.;
  Holzgrafe,~J.; Ghosh,~S.; Shams-Ansari,~A.; Puma,~E.; Sinclair,~N.;
  Reimer,~C.; Zhang,~M.; Lon\v{c}ar,~M. \emph{Adv. Opt. Photon.} \textbf{2021},
  \emph{13}, 242--352\relax
\mciteBstWouldAddEndPuncttrue
\mciteSetBstMidEndSepPunct{\mcitedefaultmidpunct}
{\mcitedefaultendpunct}{\mcitedefaultseppunct}\relax
\EndOfBibitem
\bibitem[Zheng and Chen(2021)Zheng, and
  Chen]{doi:10.1080/23746149.2021.1889402}
Zheng,~Y.; Chen,~X. \emph{Advances in Physics: X} \textbf{2021}, \emph{6},
  1889402\relax
\mciteBstWouldAddEndPuncttrue
\mciteSetBstMidEndSepPunct{\mcitedefaultmidpunct}
{\mcitedefaultendpunct}{\mcitedefaultseppunct}\relax
\EndOfBibitem
\bibitem[Hao \latin{et~al.}(2020)Hao, Zhang, Mao, Gao, Gao, Gao, Bo, Zhang, and
  Xu]{Hao:20}
Hao,~Z.; Zhang,~L.; Mao,~W.; Gao,~A.; Gao,~X.; Gao,~F.; Bo,~F.; Zhang,~G.;
  Xu,~J. \emph{Photon. Res.} \textbf{2020}, \emph{8}, 311--317\relax
\mciteBstWouldAddEndPuncttrue
\mciteSetBstMidEndSepPunct{\mcitedefaultmidpunct}
{\mcitedefaultendpunct}{\mcitedefaultseppunct}\relax
\EndOfBibitem
\bibitem[Gainutdinov \latin{et~al.}(2015)Gainutdinov, Volk, and
  Zhang]{Gainutdinov}
Gainutdinov,~R.~V.; Volk,~T.~R.; Zhang,~H.~H. \emph{Applied Physics Letters}
  \textbf{2015}, \emph{107}, 162903\relax
\mciteBstWouldAddEndPuncttrue
\mciteSetBstMidEndSepPunct{\mcitedefaultmidpunct}
{\mcitedefaultendpunct}{\mcitedefaultseppunct}\relax
\EndOfBibitem
\bibitem[Li \latin{et~al.}(2006)Li, Terabe, Hatano, and Kitamura]{LI:2006}
Li,~X.; Terabe,~K.; Hatano,~H.; Kitamura,~K. \emph{Journal of Crystal Growth}
  \textbf{2006}, \emph{292}, 324 -- 327, The third Asian Conference on Crystal
  Growth and Crystal Technology\relax
\mciteBstWouldAddEndPuncttrue
\mciteSetBstMidEndSepPunct{\mcitedefaultmidpunct}
{\mcitedefaultendpunct}{\mcitedefaultseppunct}\relax
\EndOfBibitem
\bibitem[Chezganov \latin{et~al.}(2020)Chezganov, Vlasov, Pashnina, Turygin,
  Nuraeva, and Shur]{shur}
Chezganov,~D.~S.; Vlasov,~E.~O.; Pashnina,~E.~A.; Turygin,~A.~P.;
  Nuraeva,~A.~S.; Shur,~V.~Y. \emph{Ferroelectrics} \textbf{2020}, \emph{559},
  66--76\relax
\mciteBstWouldAddEndPuncttrue
\mciteSetBstMidEndSepPunct{\mcitedefaultmidpunct}
{\mcitedefaultendpunct}{\mcitedefaultseppunct}\relax
\EndOfBibitem
\bibitem[Kang \latin{et~al.}(1987)Kang, Ding, Burns, and Melinger]{Kang}
Kang,~J.~U.; Ding,~Y.~J.; Burns,~W.~K.; Melinger,~J.~S. \emph{Opt. Lett.}
  \textbf{1987}, \emph{22}, 862--864\relax
\mciteBstWouldAddEndPuncttrue
\mciteSetBstMidEndSepPunct{\mcitedefaultmidpunct}
{\mcitedefaultendpunct}{\mcitedefaultseppunct}\relax
\EndOfBibitem
\bibitem[Luo \latin{et~al.}(2020)Luo, Ansari, Massaro, Santandrea, Eigner,
  Ricken, Herrmann, and Silberhorn]{Luo:20}
Luo,~K.-H.; Ansari,~V.; Massaro,~M.; Santandrea,~M.; Eigner,~C.; Ricken,~R.;
  Herrmann,~H.; Silberhorn,~C. \emph{Opt. Express} \textbf{2020}, \emph{28},
  3215--3225\relax
\mciteBstWouldAddEndPuncttrue
\mciteSetBstMidEndSepPunct{\mcitedefaultmidpunct}
{\mcitedefaultendpunct}{\mcitedefaultseppunct}\relax
\EndOfBibitem
\bibitem[Krasnokutska \latin{et~al.}(2019)Krasnokutska, Tambasco, and
  Peruzzo]{Krasnokutska:s}
Krasnokutska,~I.; Tambasco,~J.-L.~J.; Peruzzo,~A. \emph{Opt. Express}
  \textbf{2019}, \emph{27}, 16578--16585\relax
\mciteBstWouldAddEndPuncttrue
\mciteSetBstMidEndSepPunct{\mcitedefaultmidpunct}
{\mcitedefaultendpunct}{\mcitedefaultseppunct}\relax
\EndOfBibitem
\end{mcitethebibliography}
\providecommand{\latin}[1]{#1}
\makeatletter
\providecommand{\doi}
  {\begingroup\let\do\@makeother\dospecials
  \catcode`\{=1 \catcode`\}=2 \doi@aux}
\providecommand{\doi@aux}[1]{\endgroup\texttt{#1}}
\makeatother
\providecommand*\mcitethebibliography{\thebibliography}
\csname @ifundefined\endcsname{endmcitethebibliography}
  {\let\endmcitethebibliography\endthebibliography}{}

\end{document}